\documentstyle[aps,
prl]{revtex}
\begin{document}
\newcommand{\fe}{Sm~C$^*$~}
\newcommand{\afe}{Sm~C$^*_{A}$~}
\newcommand{\filow}{Sm~C$^*_{FI1}$~}
\newcommand{\fihigh}{Sm~C$^*_{FI2}$~}
\newcommand{\alfa}{Sm~C$^*_{\alpha}$~}
\newcommand{\feri}{Sm~C$^*_{\gamma}$~}
\draft
\twocolumn[\hsize\textwidth\columnwidth\hsize\csname@twocolumnfalse\endcsname
\title{Flexoelectricity and piezoelectricity - reason for rich
variety of phases in antiferroelectric liquid crystals}
\author{$^{+\dagger}$Mojca \v Cepi\v c and $^{+\diamond}$Bo\v stjan \v Zek\v s}
\address{$^+$J. Stefan Institute, Jamova 39, 1000 Ljubljana, Slovenia\\
$^\dagger$Faculty of Education, Kardeljeva pl. 16, 1000 Ljubljana,
Slovenia\\ $^\diamond$Institute of Biophysics, Medical Faculty,
Lipi\v ceva 2, 1000 Ljubljana, Slovenia}
\date{\today}
\maketitle
\begin{abstract}
\widetext {The free energy of antiferroelectric liquid crystal
which takes into account polar order
explicitly is presented. Steric, van der Waals, piezoelectric and
flexoelectric interactions to the nearest layers and dipolar electrostatic
interactions to the nearest and to the next
nearest layers induce indirect tilt interactions with chiral and
achiral properties, which extend to the third and to the fourth nearest
layers. Chiral indirect interactions between tilts can be large
and induce helicoidal modulations even in systems with negligible
chiral van der Waals interactions. If indirect chiral interactions
compete with chiral van der Waals interactions, the helix
unwinding is possible. Although strength of microscopic interactions change monotonically
with decreasing temperature, effective interlayer interactions change nonmonotonically
and give rise to nonmonotouous change of modulation period through various phases.
Increased enatiomeric excess i.e. increased chirality changes the phase
sequence.}
\end{abstract}
 ]
\narrowtext
Chiral polar smectics
are materials which form layered phases. In these phases the
average direction of elongated molecules is tilted for an angle
with respect to the layer normal. Smectic layers are spontaneously
polarized in the plane of the layer and perpendicular to the
direction of the tilt \cite{photinos}. Among these polar smectics
the most widely experimentally and theoretically studied materials
are ferroelectric liquid crystals. Ten years ago at an attempt to
synthesise ferroelectric liquid crystals with larger polarization,
antiferroelectric liquid crystals were discovered
\cite{takezoe89}. Their synthesis stimulated extensive
experimental research which resulted in a discovery of many new
polar liquid crystalline phases. In addition to the the
ferroelectric \fe phase and the antiferroelectric \afe phase,  the
\alfa phase and the \feri phase were found to exist in some
materials while in others instead of the \feri phase, two phases -
the \fihigh phase and the \filow phase are present. The phases are
modulated along layer normal and the modulation period of these
phases was determined by resonant X-ray scattering \cite{mach}.
The period was shown to vary nonmonotonically with decreasing
temperature from a few layers in the \alfa phase, to a few hundred
layers in the \fe phase, to approximately four layers in the
\fihigh phase, to approximately three layers in the \filow phase
and to approximately two layers in the \afe phase at low
temperatures.

Antiferroelectric liquid crystals can be theoretically described
by continuous models \cite{orihara,zeks,lorman} that predict only
two of experimentally observed phases, the ferroelectric \fe phase
and the antiferroelectric \afe phase. Discrete models
\cite{cepic95,skarabot98,madhusudana} take interlayer interactions
between the tilt vectors explicitly into account. The model
\cite{cepic95} considers competing interactions between tilts in
nearest layers and in next nearest layers and  predicts
various structures where the direction of the tilt varies
uniformly from layer to layer. The model was later called the clock
model \cite{mach}.

 Within this model the observed nonmonotonous temperature
dependence of the modulation period can be reproduced only with
nonmonotonous temperature dependences of model parameters, which
cannot be understood from microscopic interactions. Although the
model can predict structures of the \fihigh and of the \filow for
the specific set of clock model parameters, these phases cannot
exist as stable phases in a broader temperature interval. For this
some synclinic interactions would be needed between the third  and
between the fourth nearest layers \cite{madhusudana}.

In this letter we present the model where the polarization of
smectic layers is taken explicitly  into account and is treated as
a secondary order parameter to the tilt. In addition to the steric
and van der Waals interactions to the nearest layers between the
tilt vectors we consider also piezoelectric and flexoelectric
couplings between the tilt and the polarization and electrostatic
interactions between the polarizations of the nearest layers and
polarizations with the next nearest layers \cite{cepic96}.

To describe the structure of phases, we introduce the tilt vector
$\bbox{\xi}_j$ which gives the magnitude and the direction of the
tilt in the $j$-th smectic layer and the polar order parameter
$\bbox{\eta_j} = \{ \eta_{jx}, \eta_{jy} \}$ which gives the
magnitude and the direction of the in plane transverse polar order
in the $j$-th layer. The spontaneous polarization in the j-th
layer is proportional to the polar order $\bbox{\eta}$: $P_j =
P_0\;
\bbox{\eta}_j$, where $P_0$ is a layer polarization for a
completely polarly ordered layer
\cite{photinos}. Therefore  polar order parameter $\bbox{\eta}$
will be here shortly called {\it polarization}. The free energy of
the smectic system with N layers can be expressed in terms of both
order parameters as:
\begin{eqnarray}
\tilde{G} &=& \sum_{j=1}^N \frac{1}{a}\;G_0+
 \frac{1}{2}\; a_1 \; \left(\bbox{\xi}_j
\cdot\bbox{\xi}_{j+1} \right ) + \frac{1}{2}\; f_1\; \left
(\bbox{\xi}_j \times \bbox{\xi}_{j+1} \right )_z + \nonumber \\
&+& \frac{1}{2}\; b_0 \;\bbox{\eta}_j^2 + \frac{1}{2}\; b_1 \;
\left (\bbox{\eta}_j \cdot \bbox{\eta}_{j+1}\right ) + \frac{1}{8}
\; b_2 \; \left (\bbox{\eta}_j \cdot\bbox{\eta}_{j+2} \right ) +
\nonumber
\\ &+& c_p \left(\bbox{\eta}_j \times \bbox{\xi}_j\right )_z +
\frac{1}{2}\;\mu\; \left ( \bbox{\xi}_{j+1}  -  \bbox{\xi}_{j-1}
\right ) \cdot \bbox{\eta}_j.
\label{freevec}
\end{eqnarray}
Here $G_0 = 1/2\; a(T-T_0)\;\xi^2_j + 1/4\; b \; \xi^4_j $ is the tilt
dependent free energy of an isolated layer, where a continuous
transition to the tilted phase takes place at $T_0$. Dividing all
terms by the parameter $a$, we express all model parameters as
well as the free energy $\tilde{G}$ in the units of temperature.
Parameter $a_1$ corresponds to steric and van der Waals
interactions between nearest layers. Steric interactions favor
synclinic molecular alignment in neighboring layers and give a
negative contribution to $a_1$, while van der Waals intermolecular
interactions favor anticlinic ordering in neighboring layers and
give  a positive contribution to $a_1$. The parameter $a_1$ is of
the order of a few Kelvins \cite{skarabot98,link}. We believe that
steric and van der Waals interactions are negligible further than
to the nearest layers. The parameter $f_1$ gives the chiral part
of van der Waals interactions between molecules\cite{lubensky} in
neighboring layers and vanishes in racemic mixtures. We assume
that $f_1$ is smaller than $a_1$ and it is a few tenths of a
Kelvin. Polar vectors are associated with the layer polarization
via magnitude and direction of molecular dipole moments.
The parameter $b_0$ is positive because polarization is not a
proper order parameter and both $b_1$ and $b_2$ favor antiparallel
dipolar order and are therefore positive \cite{cepic96}. Since
they give intralayer and interlayer interactions which decrease for more
distant layers we expect that $b_0>>b_1>>b_2$. Transition
temperatures from the \fe phase to the  \afe phase are few degrees
lower in nonpolar racemic mixtures than in pure samples
\cite{takezoe89} which suggests that $b_0$ is also of the order of
a few K. Polarization is induced in chiral polar smectics by the
tilt and the piezoelectric coupling is given by $c_p$. The
parameter $c_p$ has also a chiral character and  vanishes in
racemic mixtures. The magnitude of the layer polarization is only
one tenth of the polarization of completely polarly ordered system
\cite{mbz} i.e. the magnitude of $\bbox{\eta}_j$ is of the order
0.1 which means that also the value of $c_p$ is of the order of a
Kelvin. The polarization coupling with the tilt in neighboring
layers is given by $\mu$ and is of the same order as $c_p$
\cite{dolganov,link99}.

The polar part of the free energy Eq.~(\ref{freevec}),
can be written in the form
\begin{equation}
G_P =\bbox{\eta} \cdot \underline{C} \cdot \bbox{\xi} + \frac 12\;
\bbox{\eta} \cdot \underline{B} \cdot \bbox{\eta}.
\label{polar}
\end{equation}
The tilt vector is $2N$-dimensional vector of
the form
$\bbox{\xi}=\{\xi_{1x},...\xi_{jx},...\xi_{Nx},\xi_{1y,}...\xi_{Ny}\}$
 and similar for the polarization $\bbox{\eta}$.
The five-diagonal $2N$-dimensional
matrix $\underline{B}$ gives intralayer and interlayer
electrostatic interactions with elements $B_{j,j}=b_0, B_{j,j\pm
1}=\frac{1}{2}\; b_1$ and $ B_{j,j\pm 2}= \frac{1}{8} \;b_2$.  The
$2N$-dimensional matrix $\underline{C}$ gives the coupling between
tilts and polarizations. The offdiagonal elements $C_{j,j\pm
1}=C_{j+N,j+N\pm 1}=\frac{1}{2}\;\mu$ and $ C_{j,j+N}=-C_{j+N,j}=
c_p$. Minimization of Eq.~(\ref{polar}) with respect to polarization
$\bbox{\eta}$ gives
\begin{equation}
\bbox{\eta}= -\underline{B}^{-1} \cdot \underline{C} \;\bbox{\xi},
\label{elimin}
\end{equation}
where $\underline{B}^{-1}$ is the inverse matrix of the
five-diagonal matrix $\underline{B}$. As
$b_0>>b_1>>b_2$ we keep in $\underline{B}^{-1}$ only the terms
up to the second order in $b_1/b_0$ and up to the first order in
$b_2/b_0$. The elements of $\underline{B}^{-1}$ are $B_{j,j}^{-1}=
\frac{1}{b_0} \left ( 1 + \frac{1}{2}\; \left(\frac{b_1}{b_0}
\right)^2 \right), B_{j,j\pm 1}^{-1}= -\frac{1}{2b_0}\;
\left(\frac{b_1}{b_0} \right)$ and $B_{j,j\pm 2}^{-1}= \frac{1}{8
b_0}\; \left ( 2\left(\frac{b_1}{b_0} \right)^2-
\left(\frac{b_2}{b_0} \right) \right)$.
 Inserting the tilt dependent polarization
Eq.~(\ref{elimin}) into
 Eq.~(\ref{polar}) we obtain the polar part of free energy as
\begin{equation}
G_P= - \frac{1}{2}\;\bbox{\xi}\; \underline{C}\;
\underline{B}^{-1}\;
\underline{C}\; \bbox{\xi}
\label{mnogo}
\end{equation}
and the free energy due to interlayer interactions is
\begin{equation}
\tilde{G}_{\mbox{int}} =
\frac 1 2\;\sum_{j=1}^N
\left ( \sum_{k=1}^4\tilde{a}_k\; \left (
\xi_j \cdot \xi_{j+k} \right )
+ \sum_{k=1}^3 \tilde{f}_k \;
\left ( \xi_j \times \xi_{j+k}
\right ) \right ).
\end{equation}
Parameters $\tilde{a}_k$ and $\tilde{f}_k$ which appear after
polarization elimination, give effective interactions between tilts.
Although direct van der Waals and steric
interactions are significant only to the nearest layers and
electrostatic interactions are significant up to the next nearest
layers, effective interactions are significant up to the fourth nearest layers.
\begin{eqnarray}
\tilde{a}_1 &=&
a_{1} +
\left (  \frac{c_p^2}{b_0}+ \frac{1}{4} \;\frac{\mu^2}{b_0}
\right )\; \left (\frac{b_1}{b_0} \right)
\nonumber\\
\tilde{a}_2 &=& \frac{1}{2}\; \frac{ c_p^2}{b_0} \left(
\frac{1}{2}
\left (\frac{b_2}{ b_0} \right ) -
\left ( \frac{b_1}{ b_0} \right )^2\right ) +
 \frac{1}{2}\;\frac{ \mu^2}{b_0}\left (1+\frac{1}{4}
 \left ( \frac{b_2}{b_0}\right) \right )
\nonumber\\
\tilde{a}_3 &=&
-\frac{1}{4}\;\frac{\mu^2}{b_0}\;\left( \frac{b_{1}}{b_{0}}\right)
\nonumber\\
\tilde{a}_4 &=& \frac{1}{8}\;\frac{\mu^2}{b_0} \;
\left (
\left(\frac{b_{1}}{b_{0}}\right )^2 -
\frac{1}{2}\;\left( \frac{b_{2}}{b_{0}}\right)
\right)
\nonumber\\
\tilde{f}_1 &=& f_1 -
\frac{2\; c_p\; \mu}{b_0}\;
\left ( 1 + \frac 1 4 \; \left(\frac{b_1}{b_0} \right )^2 +
\frac 1 8 \; \left( \frac{b_2}{b_0}\right ) \right )
\nonumber \\
\tilde{f}_2&=& \frac{c_p\; \mu}{b_0}
\; \left(\frac{b_1}{b_0}\right)
\nonumber\\
\tilde{f}_3&=& \frac 1 2 \; \frac{c_p\; \mu}{ b_{0}} \;
\left ( -\left (\frac{b_1}{b_0}\right )^2+
\frac{1}{2}\; \left(\frac{b_2}{b_0}\right)\right).
\end{eqnarray}
Achiral effective interactions between nearest layers
$\tilde{a}_1$ consist of direct steric and van der Waals
interactions ($a_1$) as well as of indirect interactions due to
piezoelectrically and flexoelectrically induced polarization.
Achiral interaction between next nearest layers are only indirect
and can be for systems with negligible flexoelectric interaction
($\mu \approx 0$) either competing \cite{cepic95}
($\tilde{a}_2>0$) or noncompeting with
($\tilde{a}_2<0$)\cite{PRL}. The second case leads to the validity
of the continuous bilayer models \cite{orihara,zeks,lorman},
although it seems that the
present experimental knowledge supports this possibility by a
single experimental evidence \cite{PRL}. In systems with large
flexoelectric interactions ($|c_p|\;\frac{b_1}{b_0} < |\mu|$),
$\tilde{a}_2$ is always positive and competes with interactions
between nearest layers $\tilde{a}_1$. Indirect effective
interactions between third neighbors are given by $\tilde{a}_3$.
The term is always negative and favors synclinic tilt directions
in interacting layers. Therefore this term in systems with
significant flexoelectric interactions tends to stabilize the
structures with three layer periodicity. Indirect
effective interactions between fourth nearest layers $\tilde{a}_4$
can be either negative or positive.

The effective chiral coupling between neighboring layers has two
contributions: the van der Waals originated ($f_1$) and the
polarization part which appears  due to the combination of piezo
and flexoelectricity ($c_p\; \mu/b_0$). Even systems with negligible
direct chiral interactions $f_1\approx 0$ are  hellicoidally
modulated. In all previous models, chiral interactions have always
been treated as weak interactions which induce only slight
perturbations in the structure i.e. helicoidal modulations with
very long periods. In antiferroelectric liquid crystals
this is not always the case. The ratio $c_p\; \mu / b_0$ can have
the value of a few Kelvins and is comparable to other
achiral interlayer interactions. The competition
between the van der Waals part of chiral interactions ($f_1$) and
indirect chiral interactions due to piezoelectric and
flexoelectric effect ($c_p\; \mu/b_0$) can explain the helix unwinding
without polarization reversal as observed in the antiferroelectric phase
of various systems \cite{takezoe89,PRL}.

The simplest solution, which minimizes the free energy
Eq.~(\ref{mnogo}) for competing interlayer interactions
($\tilde{a}_2>0$), is the clock model solution with the constant
phase difference between neighboring layers over the whole sample
\cite{cepic95}. The tilt in the j-th layer is
$
\xi_j = \theta \{ \cos (j \alpha), \sin (j \alpha) \}
$
where $\theta$ is constant and the phase difference $\alpha$ is
the difference in directions of the tilt vectors in neighboring
layers. The phase difference $\alpha$ is obtained by the
minimization of the free energy Eq.~(\ref{mnogo}) with respect to
$\alpha$.

We have analyzed the temperature dependence of solutions
for the following behavior of
model parameters with decreasing temperature: $a_1$
increases monotonically from the negative
value to the positive value, piezoelectric parameter $c_p$
monotonically increases and flexoelectric parameter $\mu$
monotonically decreases. Let us consider a microscopic mechanism
of the described parameter dependence with
decreasing temperature.  The monotonous increase
of direct interactions $a_1$ with decreasing temperature
from negative value which is due to
interpenetrating molecules through nearest layers, to the positive value in the region
where van der Waals interactions favor anticlinic order, is due to increasing smectic
order \cite{ewa}. The flexoelectric coupling  ($\mu$) decreases monotonically with
decreasing temperature, since the smectic order and hindrance of
the rotation becomes less affected by the molecules above and
below the interacting layer. In contrast, with decreasing
temperature piezoelectric coupling ($c_p$) monotonically increases since higher smectic
order strengthen the rotation hindrance within the layer.

For positive $\tilde{a}_2$ we can introduce the measure of
the competition - the {\it competition ratio}
$4 \tilde{a}_1/ \tilde{a}_2$. In racemic mixtures in
the temperature region where the competition ratio is smaller than -1, the
synclinic nonmodulated \fe phase is stable.  In
the temperature region where the competition ratio is larger than 1, the anticlinic
nonmodulated \afe phase exists. In the region
in between, the phase angle $\alpha$ changes with decreasing temperature
 rapidly from $0^\circ$ to
$180^\circ$ i.e. modulation period changes
 from infinity to two layers. In slightly chiral systems
the \alfa phase appears above the ferroelectric \fe phase.
 With additional increase of the enantiomeric excess or
increased chirality of the system (larger $c_p$), the ferroelectric \fe phase
dissapear, leaving only modulated phase with short pitch (Fig.~1).

Experimental observations have shown that in racemic mixtures only
the synclinic \fe phase and the anticlinic \afe phase exist with
the first order transition between them \cite{takezoe89}. The
narrow temperature range of the modulated phase (Fig.~1 - solid
line), can be experimentally seen as the first order transition.
In chiral samples two additional phases the \alfa phase and the \feri
phase appear. In chiral systems where piezoelectric coupling $c_p$
is important, below the transitions temperature
to the tilted phase, the modulation period is much shorter and
increases with decreasing temperature. The temperature region
where
competition ratio is between -1 and +1
becomes wider and can be experimentally observed as a
distinguished \feri phase.  The similar behavior is observed
in chiral samples of MHPOBC\cite{takezoe89} and 10OTBBB1M7 \cite{mach}
(Fig.~1 - dashed line). If piezoelectric $c_p$ coupling is still
stronger, the ferroelectric \fe phase dissapears, which was observed
in some systems\cite{joerg} (Fig.~1 - dotted line).
The present form of the  model takes into account only
quadratic interactions and can not account for first order transitions
between phases with different values of $\alpha$.

The symmetry of structures with various $\alpha$ is the
same, since the symmetry operation common to all structures
consists from the translation for a layer thickness and the
rotation for an angle $\alpha$. If $\alpha$ changes
discontinuously,
the temperature range of
different phases can be defined with DSC or similar measurements.
If transitions are continuous, the
temperature ranges of various phases are defined by changes of
the macroscopic properties and are to some extent arbitrary. The
phase difference $\alpha$ can be used as a parameter which
distinguishes between different phases. In the \alfa phase varies
from $\alpha \approx 10^\circ$ to $70^\circ$ which means that the
modulation period varies from 40 to 5 layers. In the \fe phase is
$\alpha \approx 0$ i.e. the infinite modulation period in racemic
mixtures and a few hundred layers in pure samples. In
 the \fihigh phase is $\alpha \approx 90^\circ$ or four layers, in the \filow
phase is $\alpha \approx 120^\circ$ or three layers. In the \afe
phase is $\alpha \approx 180^\circ$. The modulation period is two
layers and can also be analyzed within a bilayer model
\cite{orihara,zeks,lorman} where it is treated as a long double helix.

To conclude, we present the phenomenological model based on
microscopic intralayer and interlayer interactions. Polarization
is induced by piezoelectric and flexoelectric effect and is taken
into account explicitly. Although tilts directly interact only
with nearest layers and polarizations interact up to the next
nearest layers, indirect achiral interactions which tend to
stabilize structures with three layer and four layer periodicities
extend to the fourth neighboring layers. Chiral piezoelectric
couplings influence also the effective achiral interactions
between nearest layers. Additionally, chiral interactions due to
the flexoelectric effect are important to the third nearest layers
and can be strong. Direct interactions between nearest layers
correspond to elastic terms in continuous models. As a consequence
of the piezoelectric and flexoelectric coupling, the corresponding
Lifshitz parameter in continuous models is renormalized. In
contrast, variation of polarization, which is considered as
electrostatic interlayer coupling, has no corresponding terms in
continuous models.
The large variety of phases in antiferroelectric liquid crystals
is a consequence of a delicate balance between various mechanisms.
 In chiral
samples all interactions change with temperature monotonically.
But the competition ratio varies nonmonotonically and therefore
also phase difference $\alpha$ i.e. modulation period changes
nonmonotonically and the
ferroelectric \fe phase with long modulation period  can be stable
between two phases (\alfa and \fihigh)
with much shorter modulation periods.

Within the phenomenological discrete model we studied only  the
simplest possible interlayer interactions and the simplest
possible clock model solutions. As a future problem we left the analysis
of the influence of the phase difference $\alpha$  on the tilt which
could induce first order transitions between phases. We also did
not consider quadrupolar coupling, which could induce quadrupolar
ordering and therefore allow for distorted clock model solutions \cite{cc}.

Authors are very grateful to Ewa Gorecka and Nata\v sa Vaupoti\v c for
many stimulating discussions. The financial support of Ministry of
Education, Science and Sport of Slovenia, grant PO-518 is acknowledged.

\begin{figure}
\caption{Temperature dependence of the phase difference $\alpha $.
Model parameters change with temperature
monotonically as: $a_1 =\;(-4.1 \;\mbox{K} - 3.2\; T) $, $c_p = x\;(0.15 \;\mbox{K} -
0.16\;
T)$, $\mu = (2.12\;\mbox{K} + 1.32\; T) $. The other values were $f
= 0 \;\mbox{K};$ $ b_0 = 2\;\mbox{K};$ $ b_1 = b_0/10; $ $b_2 = b_0/100.$
The parameter $x$ gives enantiomeric excess and has the following values: $x=0$
- solid line, $x = 0.2$ - dashed line and $x = 1$ - dotted line.}
\end{figure}

\begin{references}
\bibitem{photinos} D.J. Photinos, E.T. Samulski, Science {\bf
270}, 783 (1995).
\bibitem{takezoe89} A.D.L.Chandani, E. Gorecka, Y. Ouchi, H.
Takezoe, A. Fukuda, Jpn. J. Appl. Phys., 28, 1265 (1989).
\bibitem{mach} P. Mach, R. Pindak, A.-M. Levelut, P. Barois,
H.T. Nguyen, C.C. Huang, and L. Furenlid, Phys. Rev. Lett. 81,
1015 (1998).
\bibitem{orihara} H. Orihara, Y. Ishibashi, Jpn. J. Appl. Phys. 30, L1819
(1990).
\bibitem{zeks} B. \v Zek\v s, M. \v Cepi\v c, Liq. Cryst. {\bf
14}, 445 (1993).
\bibitem{lorman} V.L.Lorman, A.A.Bulbitch, P.Toledano, Phys. Rev.E49. 1367
(1994).
\bibitem{cepic95} M. \v Cepi\v c, B. \v Zek\v s, Mol. Cryst. Liq. Cryst.
{\bf 263}, 61 (1995).
\bibitem{skarabot98} M. \v Skarabot, M. \v Cepi\v c, B. \v Zek\v s,
R. Blinc, G. Heppke, A.V. Kytik, I. Mu\v sevi\v c, Phys. Rev. E
{\bf 58}, 575 (1998).
\bibitem{madhusudana}A. Roy, N.V.Madhusudana, Eur.Phys.J.E {\bf 1}, 319
(2000).
\bibitem{cepic96} M. \v Cepi\v c, B. \v Zek\v s, Mol. Cryst. Liq.
Cryst. {\bf 301}, 221 (1997).
\bibitem{link} D.R. Link, private communication.
\bibitem{lubensky} A.B. Harris, Randall D. Kamien, T.C. Lubensky,
Phys. Rev. Lett. 78, 1476 (1997).
\bibitem{mbz} I. Mu\v sevi\v c, R. Blinc, B. \v Zek\v s, {\it The
physics of ferroelectric and antiferroelectric liquid crystals},
World Scientific, Singapore (2000).
\bibitem{dolganov} P.O. Andreeva, V.K. Dolganov, C. Gors,
R. Fouret, E.I. Kats, Phys. Rev. E 59, 4143 (1999).
\bibitem{link99} D.R. Link, G. Natale, J.E. Maclennan, N.A.
Clark, M. Walsh, S.S. Keast, M.E. Neubert,
Phys. Rev. Lett. 83, 3665 (1999).
\bibitem{PRL} I. Mu\v sevi\v c, A. Rastegar, M. \v Cepi\v c,
B. \v Zek\v s, M. \v Copi\v c, D. Moro, G. Heppke, Phys. Rev.
Lett. 77, 1796 (1996).
\bibitem{joerg}M. \v Cepi\v c, G. Heppke, J.M. Hollidt, D.
Loetsch, D. Moro, B. \v Zek\v s, Mol. Cryst. Liq. Cryst.
{\bf 263}, 207 (1995).
\bibitem{ewa} D. Pociecha, E. Gorecka, M. \v Cepi\v c, N. Vaupoti\v c,
B. \v Zek\v s, D. Kardas and J. Mieczkowski, Phys. Rev. Lett. {\bf
86}, 3048 (2001).
\bibitem{cc} P.M. Johnson, D.A. Olson, S. Pankratz, T. Nguyen, J.
Goodby, M. Hird, C.C. Huang, Phys. Rev. Lett. 84, 4870 (2000).
\end{references}
\end{document}